\documentstyle[amssymb,aps,prb,epsfig,preprint]{revtex}
\begin{document}

%@twocolumnfalse\endcsname

\title{On the possibility of superconductivity at higher temperatures in $sp$-valent diborides}
\author{J. B. Neaton and A. Perali\cite{email}}
\address{Department of Physics and Astronomy, Rutgers University,\\
136 Frelinghuysen Road, Piscataway, New Jersey, 08854-8019}
\date{\today}
\draft
\maketitle

\begin{abstract}
Superconducting transition temperatures ($T_c$'s) of MgB$_2$,
Mg$_{1-x}$Ca$_x$B$_2$, Mg$_{1-x}$Na$_x$B$_2$, and Mg$_{1-x}$Al$_x$B$_2$ are studied
within the McMillan approximation using electronic and structural
information obtained from density functional theory within
the generalized gradient approximation. The density of states 
and $T_c$ of MgB$_2$ are both shown to be extremely sensitive to volume;
in fact the density of states around the Fermi level is found to rise 
with increasing volume because of a prominent van Hove peak.
Doping the Mg sublattice with small amounts of either 
Ca, which substantially increases the unit cell volume, or Na, which 
removes an electron from the unit cell
while likewise increasing its volume, shifts the Fermi level toward the peak
and thus both types of doping are predicted to enhance $T_c$; 
in Mg$_{1-x}$Al$_x$B$_2$, however, the combined effects 
of the additional electron and decreasing average unit cell volume are shown to
decrease $T_c$ with increasing Al concentration, consistent with recent 
experiments.
\end{abstract}

\vspace{1cm}

\pacs{}

\narrowtext

%]

\section{INTRODUCTION}

Since the recent discovery of superconductivity at $T_c=39$ K
in MgB$_2$ by Akimitsu \cite{Akimitsu}, much work has been
devoted to uncovering the underlying superconducting mechanism.  Evidence for
conventional, phonon-mediated BCS-type superconductivity \cite{Parks} is now abundant.
A boron isotope effect has been observed \cite{Budko}, and photoemission measurements
indicate the opening of a superconducting 
gap in the density of states (DOS) with a narrow coherent peak
\cite{Takahashi}. The gap seems to have predominantly $s$-wave character\cite{Chen}, and
its temperature dependence follows the BCS prediction. As shown by Raman\cite{Chen}, 
photoemission\cite{Takahashi}, and some tunneling experiments\cite{Sharoni}, 
the ratio of the gap to $T_c$ is slightly larger than the BCS value, indicating 
that MgB$_2$ is in the intermediate coupling regime.

As already suggested by some authors \cite{Kortus,Pickett}
the large $T_c$ in MgB$_2$ can been understood 
with simple, weak-coupling arguments: the density of states at the Fermi level 
is large compared with that of an $sp$-valent, nearly-free electron metal of 
comparable density ($r_s = (3 \Omega/ 4\pi)^{1/3} \simeq 1.8$ a.u., 
where $\Omega$ is the volume/electron),
and the phonons most strongly coupled to electrons around the Fermi level
are due to the vibrations of the light B atoms alone.
This simple intuition, previously used to suggest a high $T_c$ in 
metallic hydrogen \cite{Ashcroft}, 
indicates two routes for increasing $T_c$: either (i) doping the B sublattice 
with lighter atoms or (ii) increasing 
the density of states at the Fermi level, $N(0)$, without significantly
diminishing the coupling to the relevant lattice modes. 
Strong-coupling effects and electron-electron interactions 
can potentially diminish the transition temperature with respect to that expected from
a weak coupling approximation, and therefore both require careful consideration, which we provide here.
In this work we focus on possibility (ii). In particular, in Sec. II
we first describe our first-principles calculations of the electronic properties of
MgB$_2$ and related compounds, and then discuss the details of the McMillan
approximation to the Eliashberg equations to evaluate $T_c$.
In Secs. IIIA and IIIB we report the structural parameters, electronic band structure, 
and density of states for MgB$_2$ as a function of volume, and for 
AlB$_2$, CaB$_2$, and NaB$_2$ at equilibrium. Importantly, a 
small {\it decrease} in electron density is shown to {\it increase} $N(0)$
and $T_c$. As an example of an empirical realization of this effect, in Sec. IIIC we
consider the result of adding small amounts of either Na, Ca, or Al
to the Mg sublattice, which in the case of Na and Ca, decreases the electron density, 
or, in the case of Al, increases the electron density.  
For Mg$_{1-x}$Al$_x$B$_2$, we
find that T$_c$ declines with increasing $x$ in accord with experiments \cite{Cava,Bianconi}. 
For Mg$_{1-x}$Na$_x$B$_2$ and Mg$_{1-x}$Ca$_x$B$_2$, however, we find that $T_c$ 
could {\it increase} up to 53 K for an Na level of $x=0.2$ or up to 52 K
for a Ca level of $x=0.2$. We conclude in Sec. IV.

%% Methods
\section{METHODS}

To investigate the structural and electronic
properties of MgB$_2$ and related compounds from first
principles, we use a plane wave-implementation \cite{vasp} 
of density functional theory \cite{kohn} within
the generalized gradient approximation (GGA) \cite{gga}
plus Vanderbilt ultrasoft pseudopotentials \cite{uspp}.  
Our Mg pseudopotential treats the low-lying 2$p$ electrons explicitly as valence states. 
For all calculations we use a dense {\bf k}-point sampling
({\bf k}-meshes of 19x19x15) and a 37 Ry plane wave cutoff,
these together converging the total energy to less than a few meV/ion.
After all lattice parameters are optimized for a given volume, 
electronic densities of states are calculated.
Once the DOS is determined, we estimate $T_c$ 
using the McMillan approximation to Migdal-Eliashberg (ME)
theory\cite{Parks,McMillan,Allen}. The ME theory includes the dynamical properties
of the effective electron-electron interaction (and therefore screening of the Coulomb repulsion) 
by introducing the Coulomb pseudopotential $\mu^*$;
self-energy corrections arising from the electron-phonon interaction are also included. 
The ME theory does not, however, include vertex corrections,
since according to the Migdal theorem such terms are smaller than the self-energy 
correction by $\lambda \sqrt{\omega_0/\omega_e}$, where $\lambda$ is the effective
electron-phonon coupling ($\lambda \sim 1$), 
$\omega_0$ the average phonon frequency at equilibrium volume, and $\omega_e$
a characteristic frequency of the order of the electronic bandwidth ($\sim 10$ eV). 
In the diboride compounds considered here, the average phonon frequency is roughly 700 K,
the frequency of the strongly-coupled E$_{2g}$ mode of MgB$_2$\cite{Pickett,Liu};
with these numbers we estimate $\lambda \sqrt{\omega_0/\omega_e}\sim 0.1$
and therefore vertex corrections can be neglected.
Moreover the DOS has a weak energy dependence in a range of 
energy of order $\omega_0$ ($\sim$ 60 meV) around the Fermi level $E_F$ (see Fig. 2),
and hence a constant density of states $N(E)=N(0)$ near $E_F$ is 
a reasonable approximation. 

With these simplifications
the solution to the Eliashberg equations for $T_c$ can be approximated
by the McMillan formula (using an estimate\cite{Parks} for the Debye temperature,
$\Theta_D \simeq (4/3) \omega_0$)
\begin{equation}
T_c=0.92\omega_0\exp
\left ( -1.04\frac{1+\lambda}{\lambda-\mu^*(1+0.62\lambda)}\right),
\label{tcmc}
\end{equation}
where $\mu^*=V_cN(0)/(1+V_cN(0)\ln (\omega_e/\omega_0))$, where $V_c$ is an
approximate static Fermi surface average of Coulomb interactions. 
Using $\omega_e \simeq$ 10 eV, we obtain $\ln(\omega_e/\omega_0)\simeq 5$; 
this implies that $V_c$ is well renormalized,
and that $\mu^*$ can reach a maximum value of 0.2 for large $V_c$.
Although a more rigorous treatment of electron-electron interactions may reduce the
value of $\mu^*$ with respect to the accepted value\cite{Rice,Ashcroft2},
in what follows we take $\mu^* = 0.1$, the conventional value for $sp$-valent,
nearly-free electron metals. Changes in $V_cN(0)$ with volume will not
have a considerable effect on $\mu^*$, which, in these compounds, 
is a slowly varying function of $V_cN(0)$ near its limiting value. 
The effective electron-phonon coupling is 
\begin{equation}
\lambda = {N(0) <g^2> \over {M\omega ^2}},
\label{coupling}
\end{equation}
where $<g^2>$ is the average over the Fermi surface of the 
elements of the square of the electron-phonon matrix and $M$ is the atomic mass,
and, in this work, varies with volume through $N(0)$ and $\omega^2$.
In the following calculations of $T_c$, the DOS is determined at each volume
from first principles, and the volume dependence of the phonon
frequency is taken to be $\omega=\omega_0 (V/V_0)^{-\gamma}$, where $\gamma$ is the
Gr\"{u}neisen parameter and $V_0$ is the equilibrium volume;
the electron-phonon matrix elements and the Coulomb pseudopotential are approximated to
be volume and concentration independent.
First-principles evaluation of the volume and composition dependence 
of these matrix elements\cite{Liu2}, which will most strongly depend 
on proper treatment of the screened electron-ion interaction,
is a formidable task for alloys and beyond the scope of this work.

%% Results
\section{RESULTS}
\subsection{Compressed and expanded MgB$_2$}

Under standard conditions MgB$_2$ crystallizes in a three atom, hexagonal primitive cell;
in it, a single Mg atom is centered one half-lattice vector 
above a honeycomb boron network. The lattice parameters have been measured to be
$a=3.084$ {\AA} and $c=3.523$ {\AA} at low temperatures \cite{Akimitsu,Matko}, and
we calculate $a=3.065$ {\AA} and $c=3.519$ {\AA}, in excellent agreement 
with these experiments and also with previous calculations \cite{Loa,Vogt}. 
Detailed discussions of the electronic structure of MgB$_2$ have already appeared
in a number of recent studies \cite{Kortus,Pickett}. Two dimensional hole-like
$\sigma$-bands (arising from intraplanar B $p_{x,y}$-like orbitals) result in
a flat DOS with a prominent van Hove spike. 
A three dimensional $p_z$-like band also weakly contributes to the DOS.
Our calculated bands at equilibrium volume (Fig. 1(b)) support these views and
agree well with recent studies \cite{Kortus,Ivan,Satta,Medve}.

To examine the effects of volume on the 
electronic structure near the Fermi level, we optimize the structural parameters of
MgB$_2$ at several densities, corresponding to changes in volume of $\pm$30\%, 
and calculate the energy bands and density of states; the results appear in Figs. 1 and 2.
The $\sigma$-bands are largely nearly-free electron-like: their dispersion is parabolic
near the $\Gamma$ point, and
their overall bandwidth is comparable to the free electron value at
this density ($\sim$ 15.5 eV). In addition, the 
bandwidth increases as volume decreases, scaling roughly as $a^{-2}$ as might
be expected for free electrons, and the nearly-free electron gaps at the zone boundaries
grow with increasing volume. The weak dispersion of the $\sigma$-bands along $\Gamma$A
reflects their particular quasi-two dimensionality, decreasing with increasing volume
(and increasing $c$-axis).

The position and shape of these bands are reflected in the density of states appearing in Fig. 2,
which we find to depend strongly on the cell volume. With respect to 
its value at equilibrium, $N(0)$ decreases by as much as 24\% at the highest compression
considered. Correspondingly, as shown in Fig. 2, $N(0)$ 
increases as the lattice is expanded, contrary to expectations 
for a nearly-free electron metal.
The dependence of the DOS on volume is almost entirely due to the changes in the
width and position of a considerable van Hove peak ($\sim 2$ eV below the Fermi level), 
which originates from a saddle point in the highest occupied $\sigma$-band at the M point.
The decrease in bandwidth with increasing volume reduces the seperation between
the peak and the Fermi level, driving up the density of states.
The singularity is further enhanced by an increase in two dimensionality (due to the elongated $c$-axis
evident in Fig. 2's inset), and a complementary increase in the primary Fourier component of
the electron-ion interaction (evident from the larger gap at the zone boundary).
Because these $\sigma$-bands are essentially two dimensional, they result
in a nearly constant DOS near the top or the bottom of the bands (where the dispersion is
most free electron-like); their contribution to the DOS does
not change appreciably with volume, and indeed, 
the DOS at each expanded volume takes on nearly the
same value roughly 0.3 eV above $E_F$ (Fig. 2). 

As a result of the apparent increase in $N(0)$ for lower densities (i.e., larger
volumes), the superconducting transition temperature $T_c$ should rise, and to 
verify this explicitly we calculate the transition temperature $T_c$ using Eqs. (1) and (2).
We first determine the electron-phonon coupling by requiring that 
T$_c=39.5$ K at the equilibrium volume. Using $\omega_0=700$ K and $\mu^*=0.1$, 
we find that a coupling value of $\lambda=0.835$ is required \cite{note}, 
in agreement with recent estimates\cite{Kortus,Pickett,Liu} of $\lambda$ between 0.7 and 0.95.
We then calculate $T_c$ for a range of volumes ($V/V_0=0.7-1.3$)
using this value of $\lambda$, with the results appearing in Fig. 3.
A general trend is clear: at lower volumes (positive pressure) 
$T_c$ decreases, while as volume increases
(negative pressure) $T_c$ increases, in accordance with the DOS at the Fermi level.
The variation in $T_c$ does depend on the Gr\"uneisen parameter,
and for larger values of $\gamma$, $T_c$ can change quite substantially over
the volume range considered here. For example $T_c$ ranges from 26 to 50 K for $\gamma=0$
but for $\gamma=2$ it can reach 79 K. A larger value of $\mu^*$ will lower $T_c$:
our estimate of $T_c=$ 62 K at $V/V_0=1.1$ with $\mu^*=$0.1 and $\gamma=2$ (see Fig. 3)
drops to 51 K for $\mu^*=0.15$ (for fixed $\lambda$).
As MgB$_2$ possesses equal numbers of electron
and hole carriers, we also note that correlated electron-hole fluctuations
may reduce $\mu^*$ and increase our predicted transition temperatures\cite{Rice,Ashcroft2}, 
as mentioned above.
Hydrostatic pressure has already been found to decrease the transition 
temperature, with a slope in the range of $-1.6$ K/GPa \cite{Lorenz} and $-1.1$ K/GPa \cite{Tomita}. 
Extrapolating to low pressures from data at $V/V_0 =0.9$ ($p \simeq 19$ GPa), 
we find a slope of -1.2 K/GPa for $\gamma=1.5$, slightly lower
than a previous calculation\cite{Loa} of this slope using $\gamma=1$.
The value of $\gamma$ is not experimentally known, 
and it is used here as an adjustable parameter \cite{Seiden};
we reserve its explicit calculation from first principles for a future study.

%% Results: electronic structure of (Na,Li)B_2, Mg1-xNaxB_2, Mg1-xAlxB_2, 

\subsection{Increasing or decreasing $T_c$: the effects of doping with Ca, Na, or Al}

While the large effect of volume (and electron density) on $T_c$ is evident, 
isotropic expansion of the lattice is generally not possible experimentally.
Epitaxial growth of MgB$_2$ is a one possible solution, and recently, 
Kang {\em et al.}\cite{Kang} report the growth of high-quality MgB$_2$ 
thin films on Al$_2$O$_3$ substrates having $T_c=39$ K with
a sharp transition width of $0.7$ K. Using different substrates, structures with a larger volume
might be realized, though with different $c/a$ ratios. 
A more direct method would be to replace Mg with a larger ion. Calcium,
with an ionic radius\cite{Kittel} of 0.99 {\AA} (compared to 0.65 {\AA} for Mg),
would certainly increase the volume of the unit cell, 
as verified by recent first principles calculations \cite{Medve}.
On intuitive grounds, sodium would be a particularly attractive dopant; 
as suggested previously (see, e.g., Refs. \cite{Kortus,Medve,Suz})
it carries one less valence electron and would 
therefore add holes to the system (hole doping). Additionally,
with an ionic radius of 0.97 \AA, sodium should increase the unit cell volume.
Aluminum, on the other hand, adds electrons to the system and reduces the volume, and, again 
following intuition, small amounts of Al have already been shown \cite{Cava,Bianconi} to 
decrease $T_c$.

In what follows we study the behavior of $T_c$ for small amounts of Ca, Na, and 
Al in MgB$_2$ quantitatively. Important physical changes due to the presence
of these dopants are first elucidated through calculations of the equilibrium 
structural and electronic properties of their hypothetical
bulk (MgB$_2$-like) hexagonal structures. In agreement
with expectations based on the ionic radii and previous calculations \cite{Medve,Suz},
replacing Mg with either Ca or Na {\it increases} the unit cell volume:
our calculated volumes of Ca (36.40 {\AA$^3$}) and Na (34.11 {\AA$^3$}),
are 27\% and 19\% larger than that of MgB$_2$ (28.63) {\AA$^3$}, respectively.
For comparison, we find that AlB$_2$ stabilizes with a primitive cell volume of 25.54 {\AA$^3$}, 
11\% {\it smaller} than that of MgB$_2$ and in excellent agreement 
with the experimental value\cite{Matko} of 25.58 {\AA$^3$}. 
In both NaB$_2$ and AlB$_2$ the intraplanar B-B distance
{\it decreases}, but by less than 2\%; in CaB$_2$, however, the B-B distance {\it increases}
by 5\%. Adding a larger ion increases the $c$-axis for both Na (by 22\% to 4.30 {\AA})
and Ca (15\% to 4.069 {\AA}); 
in AlB$_2$, however, the $c$-axis decreases by roughly 7\% to 3.282 {\AA}. 
Thus by doping the Mg sublattice and changing the $c$-axis,
the two dimensionality of the $\sigma$-bands
can be either enhanced or diminished, and this will also have important consequences
for the density of states near the Fermi level, as we now show.

In Fig. 4(a) we plot the DOS of  MgB$_2$ and CaB$_2$.  Mg and Ca are isovalent, but according 
to the results of the previous section, adding Ca should have a
large effect on the density of states. Indeed, CaB$_2$ would result in 
an $N(0)$ of 0.981 states/eV-cell, an increase of 41\% over that in bulk MgB$_2$.
Note the increase in magnitude of the singularity: the larger volume gives the $\sigma$-bands
more two dimensional character, and the lower electron density
results in a smaller bandwidth. Finally, we note that although the unoccupied $d$-states  
do weakly contribute to the density of states near the Fermi level through some hybridization with
the $p_z$-band, they have only a minimal effect on the $\sigma$-bands.

In Fig. 4(b) we plot the DOS of NaB$_2$, MgB$_2$, and AlB$_2$.
Because Na, Mg, and Al are heterovalent, we expect changes
both in the number of electrons per cell and in the volume 
to have a profound effect on the density of states near the Fermi level.
In NaB$_2$ the DOS at the Fermi level 
substantially increases, from 0.711 to 1.014 states/eV-cell;
in contrast, $N(0)$ drops to 0.378 states/eV-cell in AlB$_2$. 
By removing one electron, the Fermi level of NaB$_2$
is shifted downward relative to that of MgB$_2$ and toward the van Hove singularity arising from the 
two dimensional $\sigma$-bands. As discussed above, the larger volume 
again results in a more pronounced van Hove singularity just below the Fermi energy,
as in CaB$_2$. In AlB$_2$ the opposite occurs with respect to MgB$_2$; the Fermi level shifts away from a
van Hove peak that is now less prominent due to the increase in
electron density and decrease in volume.

The large increase in $N(0)$ for both CaB$_2$ and NaB$_2$ clearly raises the possibility
of a $T_c$ larger than that of MgB$_2$. Indeed for the hypothetical CaB$_2$ we
obtain $T_c = 65.6$ K and for NaB$_2$ we calculate $T_c=68.5$ K (using $\gamma=0$, and
$<g>$, $\omega_0$, and $\mu^*$ appropriate for MgB$_2$). But the expanded $c$-axis and appreciable  
DOS at the Fermi level, of course, call in to question the stability of 
the resulting CaB$_2$ or NaB$_2$ lattice. Neither compound, to our knowledge, 
has yet been synthesized. Therefore, instead of focusing on bulk CaB$_2$ or NaB$_2$,
we concentrate on the electronic properties and transition temperatures of
the alloys Mg$_{1-x}$Ca$_x$B$_2$ and Mg$_{1-x}$Na$_x$B$_2$ 
for $x$ up to 0.2. Previous calculations\cite{Medve}
using supercells constructed from the experimental lattice parameters of MgB$_2$ have been
performed for large Li and Na doping fractions of $x=0.25$ and $0.5$. 
To obtain a quantitative estimate of $T_c$ for small $x$ in the absence of experimental data and without
resorting to large supercells, we calculate $N(0)$ from first principles using a simple rigid
band model {\it that also incoporates the effects of the changing density} due
to changes in the lattice parameters. To facilitate comparison with Na and also test the 
validity of our results against existing experiments, we also consider Mg$_{1-x}$Al$_x$B$_2$.

To calculate $T_c$, an estimate for $N(0)$ is required for small $x$.
As we have shown, the DOS of CaB$_2$ departs considerably from that of MgB$_2$.
Given that it is isovalent with Mg, the foremost consequence
of adding small amounts of Ca for the density of states will be the increase in volume,
though the change in electron-ion interaction will also have an effect, e.g., 
through the contribution of a small $d$ state density near the Fermi level.
In contrast, the most important consequence
of adding small amounts of either Al or Na will certainly be a rigid shift of the Fermi level. 
But adding either Al or Na will also modify the electron density in two additional ways: first,
as these atoms are not isovalent to Mg, the number of 
electrons will change; and second, the nominal charge of the 
metallic ion in the cell will clearly differ from that of Mg.
Both will in turn affect the electron-ion
interaction. The strong volume dependence of the DOS has already been demonstrated above;
to illustrate the contribution of the electronic and ionic charge 
explicitly, in Fig. 5 we plot the band structures
of NaB$_2$, MgB$_2$, and AlB$_2$ {\it calculated with the lattice parameters of MgB$_2$}
(i.e., for {\it fixed} volume). At first glance the bands are quite similar,
differing only by a shift of the Fermi level. Small differences are present, however, 
and they result from changes in the electron-ion interaction, the 
electronic density, and the ionic charge. For example, 
in NaB$_2$ the $p_z$ band crossing $E_F$ (whose constituent orbitals
extend out of the B planes and closest to the metallic ion) shifts upward due to the weaker
attraction to the ion \cite{Pickett}, requiring the $\sigma$-bands to shift downward relative to
the Fermi level. In AlB$_2$ the attraction is stronger and 
the $p_z$ band shifts the opposite way.  Further, the gaps at the zone boundaries are largest
for Na and smallest for Al.

In the following determination of $N(0)$ we explicitly include, as mentioned above, the effects of
volume due to the dopant ion. As a working hypothesis we assume that the effects of the electron-ion
interaction will be minimal, partially justified by the observed 
small differences in the $\sigma$-bands (which are most important to the DOS) when 
Mg is substituted for Ca at fixed volume. Furthermore, for NaB$_2$ and AlB$_2$, we also 
implicitly neglect the effects of the extra electronic and ionic charge and perform all 
calculations with an Mg pseudopotential. In this context we note that 
further work on the impurity systems, using the virtual crystal or coherent potential approximations, 
would therefore be of interest. In order to include the effects of small dopant amounts
on the average unit cell volume, we extrapolate linearly from the $a$ and $c$ determined
for MgB$_2$ to those of CaB$_2$ or NaB$_2$, thus obtaining $a(x)$ and $c(x)$,
where $x$ is the doping fraction. For AlB$_2$, we make use of available experimental data,
as we explain below. We then calculate the DOS of MgB$_2$ for
a few values of $a(x)$ and $c(x)$, and, if we are considering either Al or Na,  
we adjust the Fermi energy by an amount $\Delta\mu(x)$ so that the integral of the calculated DOS is equal
to $6+2(1-x)+xZ_{d}$, where the dopant valence $Z_{d}$ is 3 for Al and 1 for Na.
(When treating small amounts of Ca, we simply take the DOS at the Fermi level of the expanded MgB$_2$.)
$T_c$ is then determined in the manner described above, 
keeping $\mu^{*}$ fixed but scaling the electron-phonon coupling (Eq. 2)
with a Gr\"uneisen parameter through the variation of $\omega^2$.  
The variation of $\mu^*$ with volume and concentration is difficult to estimate quantitatively, particularly
since although $N(0)$ increases with increasing volume, the static Coulomb interaction $V_c$ 
declines. For these reasons, however, we may expect its variation with volume
to be small compared with the electron-phonon
coupling. Finally, in all calculations we approximate the electron-phonon 
matrix elements $<g^2>$ as fixed.

\subsubsection{Mg$_{1-x}$Ca$_x$B$_2$}

Table I displays the results of our calculations of the $N(0)$ of MgB$_2$
for lattice parameters corresponding to Mg$_{1-x}$Ca$_x$B$_2$ up
to $x=0.2$. As expected, the increase
in volume again increases $N(0)$, and, as can be seen from Fig. 6, this increases
$T_c$. For $\gamma=2$, $T_c$ can reach up to 52 K for $x=0.2$, despite the fact
that the total number of electrons and holes in each cell remains fixed.
By calculating $N(0)$ with Mg ions only, we neglect changes in the electron-ion
interaction, but these should not be important to the density of states near
the Fermi level, where the $\sigma$-bands are most important.
The stability of CaB$_2$ (i.e., $x=1$) is of course questionable, given that Ca and B prefer
to equilibrate in the cubic hexaboride structure\cite{Matko}, but if it were possible to introduce
even small amounts into the MgB$_2$ lattice, an appreciable rise in $T_c$ could be
detected.

\subsubsection{Mg$_{1-x}$Na$_x$B$_2$}

In Table II we report DOS and lattice parameters 
as a function of Na doping fraction $x$ at the adjusted Fermi level of MgB$_2$, $\Delta\mu(x)$.
Note that $N(\Delta\mu(x))$ increases by 7\% for $x=0.2$ with respect to $x=0$.
This conflicts with the calculations of Medvedeva {\it et al.}\cite{Medve}, who report no increase in
DOS at the Fermi level at $x=0.25$ and only a 4\% increase at $x=0.5$ in their
supercell calculations without relaxation of the lattice parameters. Failure to treat the effects
on the DOS due to the increase in volume is the likely reason for this discrepancy.
In Fig. 7 we plot T$_c$ as a function of Na doping fraction $x$ for different Gr\"uneisen
parameters. For $\gamma=2$, a $T_c$ of $\sim$53 K would be expected
for 20\% Na.  Although we explicitly considered the important effects of
changing volume on $N(0)$, as mentioned above we neglect the effects of the change in ionic
charge and the contribution of the reduced number of electrons to the
electron density.  Conveniently these effects should be small and 
may even largely counterbalance one another, as follows.
Replacing the Mg ion with Na will shift the average position of the
$\sigma$- and $p_z$-bands with respect to the Fermi level, as discussed
above, but, as we have emphasized, $N(0)$ is most sensitive to the van Hove peak arising from
the $\sigma$-bands. Although this electrostatic effect shifts the Fermi level 
slightly {\it away} from the peak in the case of Na doping, the larger
van Hove peak due to the increased two dimensional character
should compensate. Further experiments will play a crucial role in determining the
feasibility of Na doping, but our work nonetheless
serves to illustrate the positive effects of increasing volume on the transition temperature,
even for small Na concentrations. 

\subsubsection{Mg$_{1-x}$Al$_x$B$_2$}

To examine the effects of small amounts of Al on $T_c$ we use the lattice parameters
and electron density measured by Bianconi {\it et al.} \cite{Bianconi};
we then calculate the density of states at the Fermi level and $T_c$ as above\cite{notebian},
using an Mg pseudopotential. From Table III it is clear that doping the Mg sublattice with Al decreases the DOS at
the Fermi level, and as is apparent from Fig. 8, the $T_c$'s we calculate
reproduce the data quite well using a Gr\"uneisen parameter of 1.2.
Larger (smaller) values of $\gamma$ shift the curves to lower (higher) temperatures 
but do not change the overall shape, which is dominated by the changes
in the density of states. Simply shifting the Fermi level without
changing the volume (and therefore with $\gamma=0$) results 
in a $T_c$ that overestimates the data considerably at larger concentrations
(by more than 3 K at $x=0.2$).
In the case of Al doping the secondary electrostatic effect due to the increase in ionic charge
would shift the Fermi level toward the van Hove peak,
but as the doping fraction increases, the position of this peak is less relevant, since
electrons first fill the $\sigma$-bands, shifting the Fermi level 
into an energy range dominated by the three dimensional $p_z$ band,
whose contribution to the DOS is significantly less than that of the 2D $\sigma$-bands.
Thus the contribution from the changing DOS (via rigid shift {\it and} changing
volume) is evidently most important to $T_c$ for Al concentrations up to 8\%.
The additional electrostatic consequences stemming from the increase
in ionic charge and electron density are artificially absorbed into $\gamma$, but we
expect these effects to be minimal for the reasons argued above.

%% Conclusions
\section{CONCLUSIONS}

We have found here that the superconducting transition temperature 
is quite sensitive to the unit cell volume and in particular,
increasing the volume should increase $T_c$.
For the isotropically expanded lattice, the
van Hove peak from the two dimensional bands increases
the density of states around the Fermi level.
The importance of this effect has also been demonstrated through calculations of 
three other $sp$-valent diborides, CaB$_2$, NaB$_2$, and AlB$_2$, and also
through the alloys Mg$_{1-x}$Ca$_x$B$_2$, Mg$_{1-x}$Na$_x$B$_2$, and Mg$_{1-x}$Al$_x$B$_2$.
Additional small amounts of Ca are found to {\it increase} $T_c$ up to 52 K due to the increase
in volume without valence change. Strontium or barium, other possible dopants likely
to increase the volume, would also be of interest, though low-lying $d$ states may play a 
more significant role near the Fermi level. The presence of Na impurities in small amounts is
also shown to {\it increase} $N(0)$ and $T_c$ up to 53 K, contrary to the theory
of hole superconductivity \cite{Hirsch}. 
Lithium is another possible hole dopant, but since we calculate
the equilibrium volume of LiB$_2$ to be 8.9\% less than that of MgB$_2$, the decrease in volume may
compensate for the shift in the Fermi level toward higher density of states,
resulting in only a small variation in $T_c$.
Doping with Al is found to decrease $N(0)$ and $T_c$ for small $x$, and a rigid shift of the Fermi
level of a compressed MgB$_2$ lattice results in good agreement with recent experiments.  
Finally, we hope that this work will motivate experimental studies of impurities in MgB$_2$
that decrease the average electron density.
 
\acknowledgments

We appreciate valuable discussions with 
N. W. Ashcroft, A. Bianconi, M. H. Cohen, G. Kotliar, and D. Vanderbilt. 
We thank G. Kresse and J. Hafner for providing VASP.
A. Perali acknowledges partial support from
Fondazione ``Angelo della Riccia''.

% Table 1

\begin{table}
Table I. Lattice parameters $a$ and $c/a$,
and density of states at the Fermi level $N(0)$ (states/eV-cell)
for different Ca concentrations $x$ in Mg$_{1-x}$Ca$_x$B$_2$ as calculated in this work.
\begin{tabular}{l c c c}
                $x$ & $a$ (\AA) & $c/a$ & $N(0)$ \\\hline
	        0.00    & 3.065   & 1.148   & 0.711\\
                0.05    & 3.072   & 1.154   & 0.714\\
                0.10    & 3.080   & 1.160   & 0.717\\
                0.15    & 3.087   & 1.167   & 0.723\\
		0.20    & 3.095   & 1.173   & 0.727\\
\end{tabular}
\end{table}

% Table 2

\begin{table}
Table II. Lattice parameters $a$ and $c/a$, total number of electrons/cell $N_e$, 
Fermi level shift $\Delta\mu$,
and density of states at the shifted Fermi level N($\Delta\mu$) (states/eV-cell)
for the different Na concentrations $x$ in Mg$_{1-x}$Na$_x$B$_2$ as calculated in this work.
\begin{tabular}{l c c c c c}
                $x$ & $a$ (\AA) & $c/a$ & N$_e$ & N($\Delta\mu$) & $\Delta\mu$ (eV) \\\hline
	        0.00    & 3.065 & 1.148 & 8.00 & 0.711 & 0.000\\
                0.05    & 3.063 & 1.162 & 7.95 & 0.724 & 0.069\\
                0.10    & 3.061 & 1.175 & 7.90 & 0.739 & 0.139\\
                0.20    & 3.057 & 1.203 & 7.80 & 0.760 & 0.282\\
\end{tabular}
\end{table}

% Table 3

\begin{table}
Table III. Lattice parameters $a$ and $c/a$, total number of electrons/cell $N_e$,
Fermi level shift $\Delta\mu$,
and density of states at the shifted Fermi level N($\Delta\mu$) (states/eV-cell)
for different Al concentrations $x$ in Mg$_{1-x}$Al$_x$B$_2$.
The $x$, $a$, $c/a$, and N$_e$ are taken from Bianconi {\it et al.} \cite{Bianconi};
in particular note that we use the experimental lattice parameters even in the case of $x=0$ to better
compare with experiment. N($\Delta\mu$) and $\Delta\mu$ are calculated as described in the text.
\begin{tabular}{l c c c c c}
                $x$ & $a$ (\AA) & $c/a$ & N$_e$ & N($\Delta\mu$) & $\Delta\mu$ (eV) \\\hline
	        0.00    & 3.085   & 1.141   & 8.00 &  0.720 & 0.000\\
                0.02    & 3.084   & 1.140   & 8.02 &  0.716 & 0.028\\
                0.04    & 3.083   & 1.139   & 8.04 &  0.709 & 0.063\\
                0.08    & 3.081   & 1.135   & 8.08 &  0.700 & 0.115\\
\end{tabular}
\end{table}

% Fig. 1

%@twocolumnfalse\endcsname

\begin{figure}[h]
\centering \psfig{file=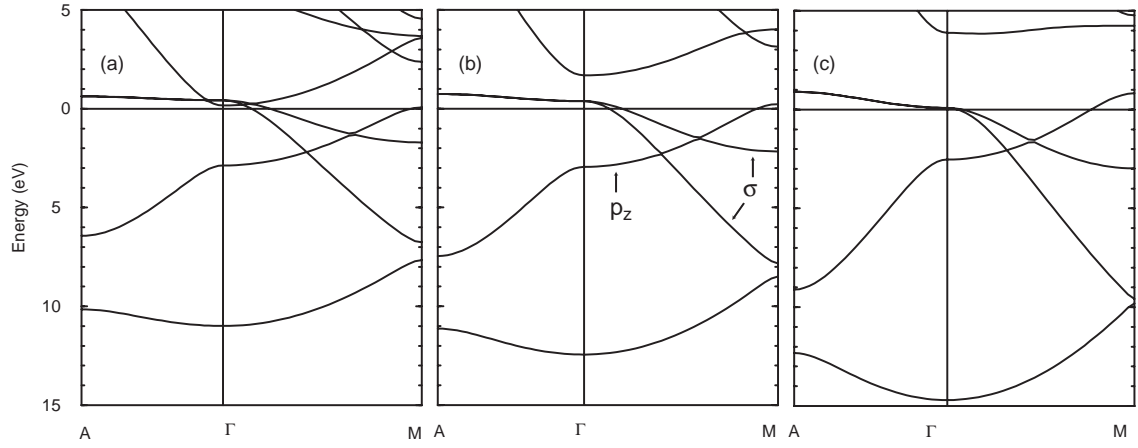, width=15cm}
\caption{Energy bands of MgB$_2$ (a) after expansion (V/V$_0$ = 1.3), (b) 
at equilibrium (V/V$_0$ = 1.0), and (c) under compression (V/V$_0$ = 0.7).
$V_0$ is the equilibrium volume, which we calculate to be 28.63 {\AA}$^3$.}
\label{bands1}
\end{figure}

%]

% Fig. 2

\begin{figure}[h]
\centering \psfig{file=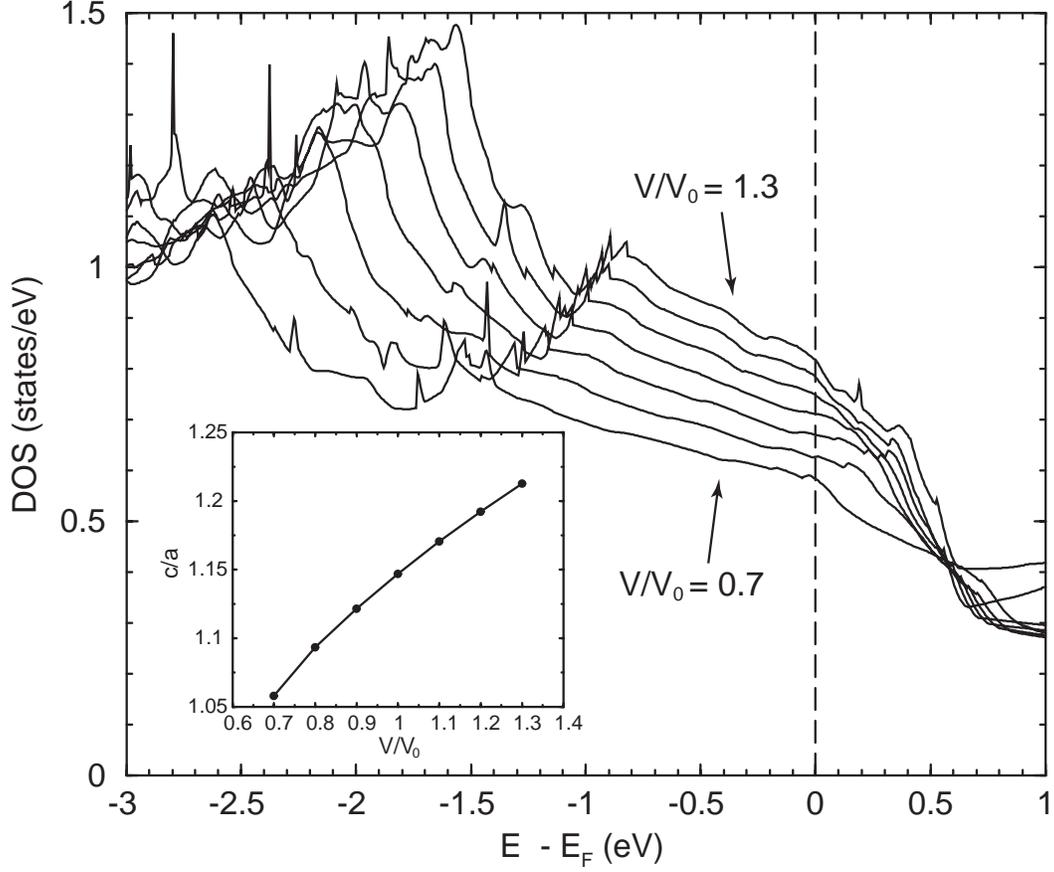, width=14cm}
\caption{Density of states of MgB$_2$ as a function of volume;
$V_0=28.63$ {\AA}$^3$ is the calculated equilibrium volume.)
The inset shows the calculated $c/a$ ratio as a function of volume.
At each volume both $a$ and $c/a$ are fully relaxed. We obtain $a=2.797$ {\AA}
and $c/a=1.0579$ for $V/V_0=0.7$ ($p\sim$ 97 GPa); and we calculate
$a=3.284$ {\AA} and $c/a=1.213$ for $V/V_0=1.3$.}
\label{dosexp}
\end{figure}

% Fig. 3

\begin{figure}[h]
\centering \psfig{file=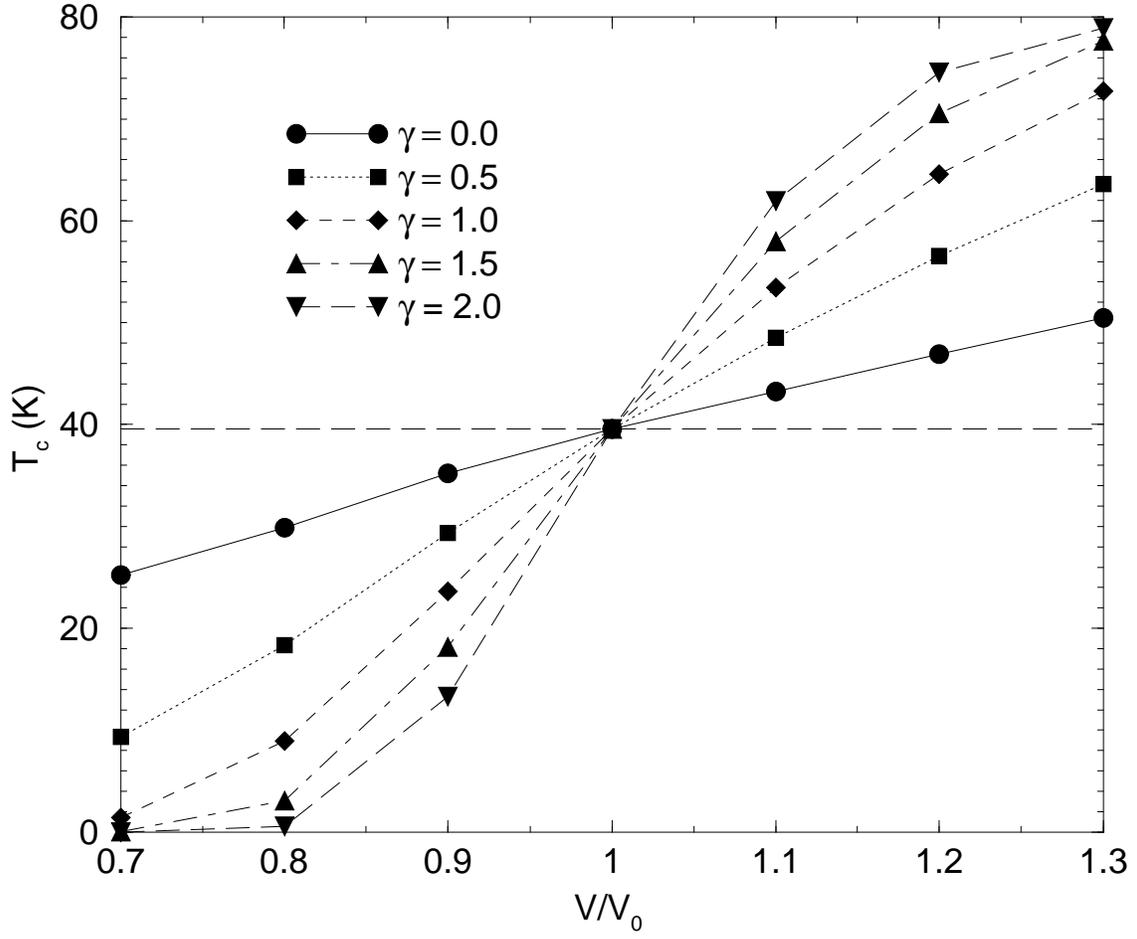, width=15cm}
\caption{Calculated transition temperatures determined from 
the McMillan equation using $\mu^*=0.1$ as a function of volume for MgB$_2$
with different Gr\"uneisen parameters $\gamma$.}
\label{tcexp}
\end{figure}

% Fig. 4

\begin{figure}[h]
\centering \psfig{file=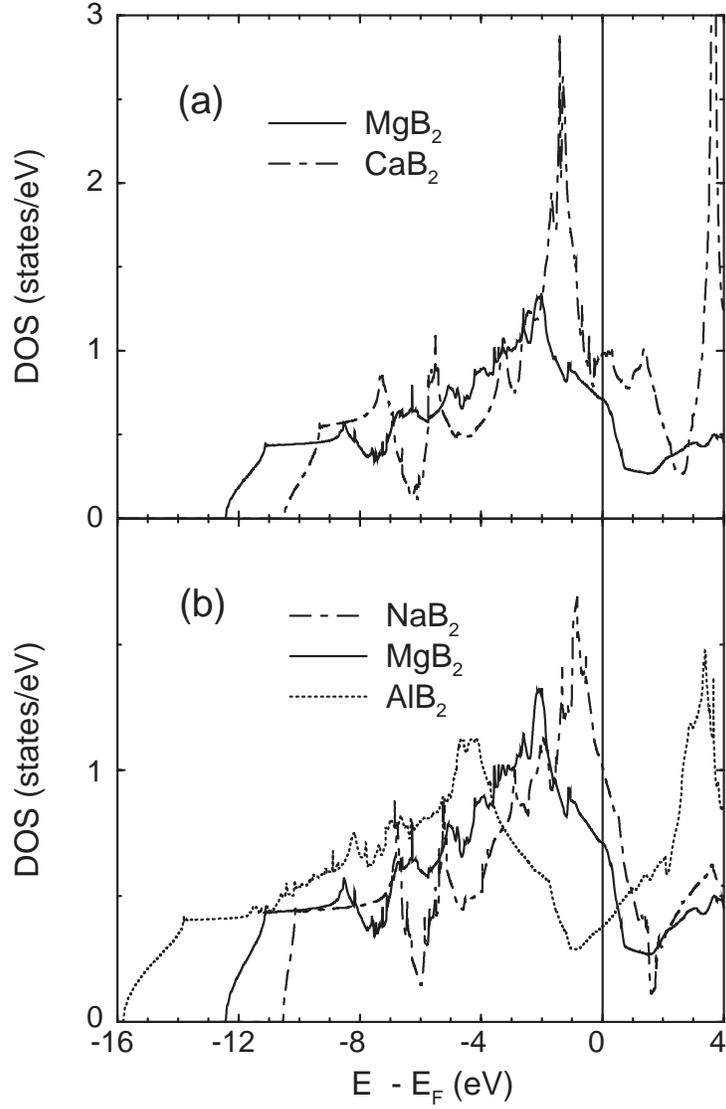, width=10cm}
%\centering \psfig{file=fig2.eps, width=10cm}
\caption{Density of states of (a) MgB$_2$ and CaB$_2$ at their
calculated equilibrium volumes. Density of
states of (b) NaB$_2$, MgB$_2$, 
and AlB$_2$, also at their calculated equilibrium volumes.}
\label{dosab2}
\end{figure}

%@twocolumnfalse\endcsname

% Fig. 5

\begin{figure}
\centering \psfig{file=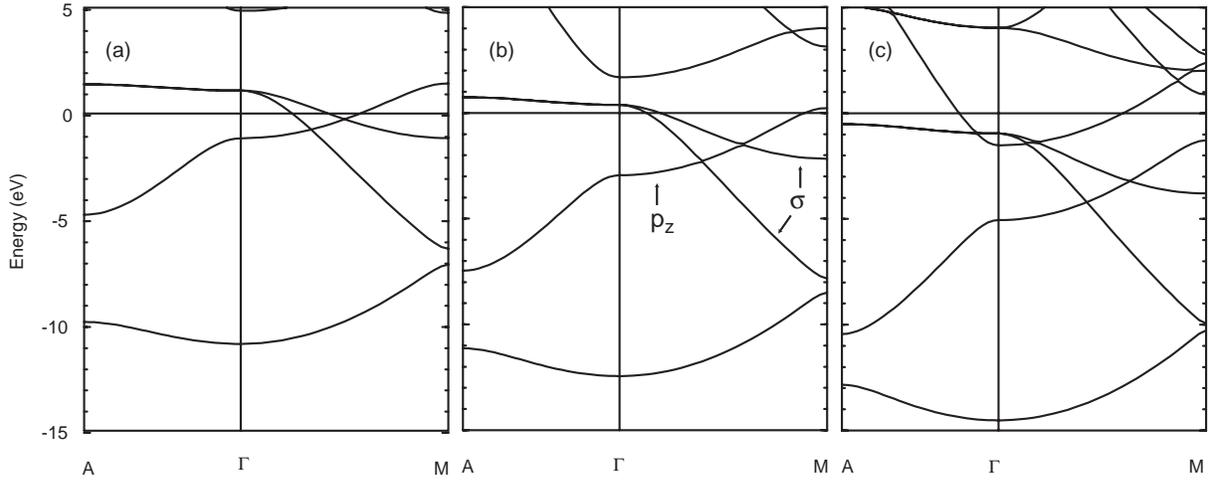, width=16cm}
\caption{Bands of (a) NaB$_2$, (b) MgB$_2$, and (c) AlB$_2$ calculated 
with the structural parameters of MgB$_2$. With the volume fixed,
the most noticable difference between bands is 
a rigid shift of the Fermi level, though additional dissimilarities are evident
from changes in the electron-ion interaction, electron density,
and ionic charge as Mg is replaced with Na or Al.}
\label{bands2}
\end{figure}

%]

% Fig. 6

\begin{figure}[h]
\centering \psfig{file=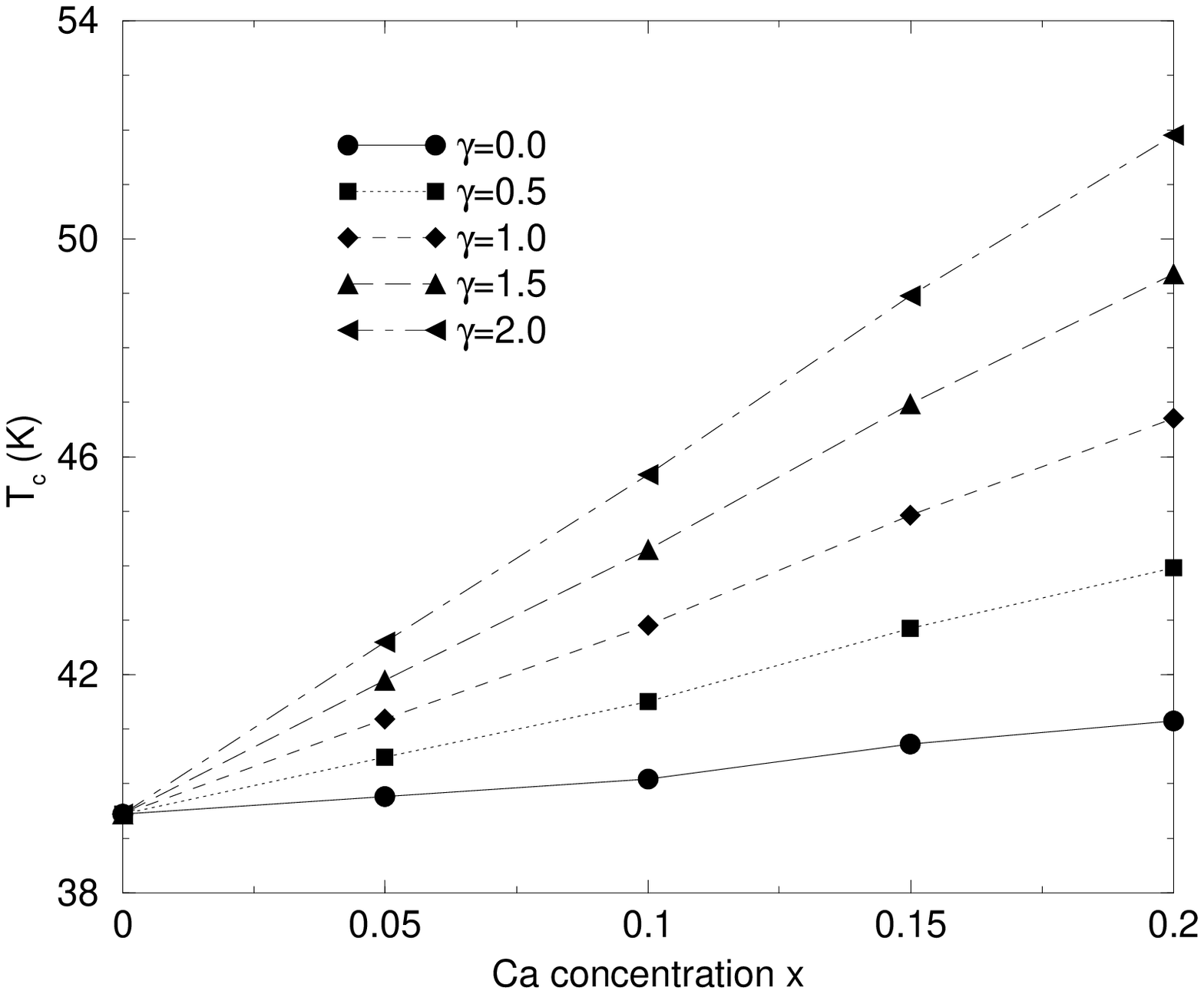, width=15cm}
\caption{Predicted transition temperatures (calculated with
the McMillan equation using $\mu^*=0.1$) as a function of Ca concentration 
$x$ for Mg$_{1-x}$Ca$_x$B$_2$ with different Gr\"uneisen parameters $\gamma$.}
\label{tcab8}
\end{figure}

% Fig. 7

\begin{figure}[h]
\centering \psfig{file=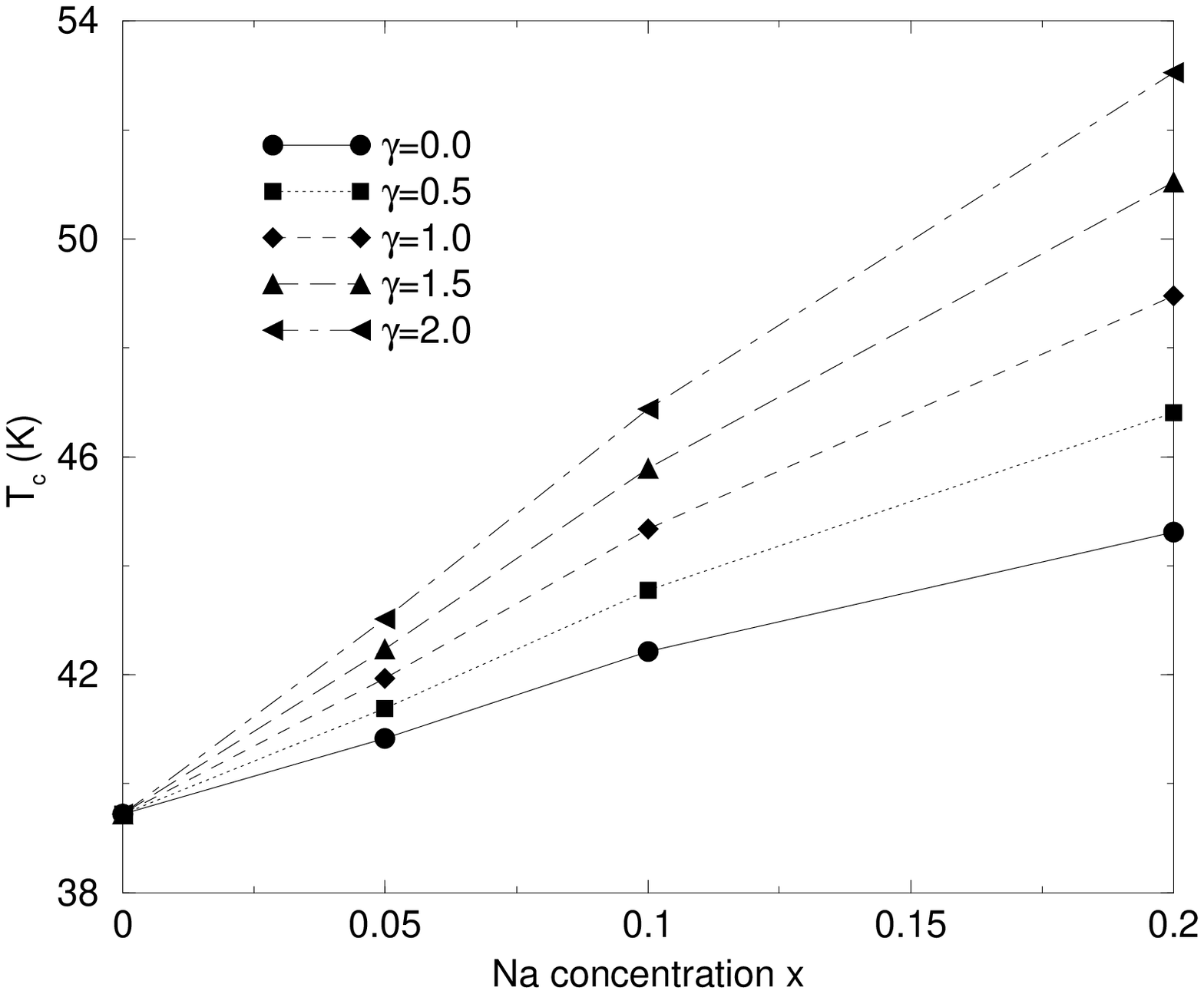, width=15cm}
\caption{Predicted transition temperatures (calculated with
the McMillan equation using $\mu^*=0.1$) as a function of Na concentration 
$x$ for Mg$_{1-x}$Na$_x$B$_2$ with different Gr\"uneisen parameters $\gamma$.}
\label{tcab2}
\end{figure}

% Fig. 8

\begin{figure}[h]
\centering \psfig{file=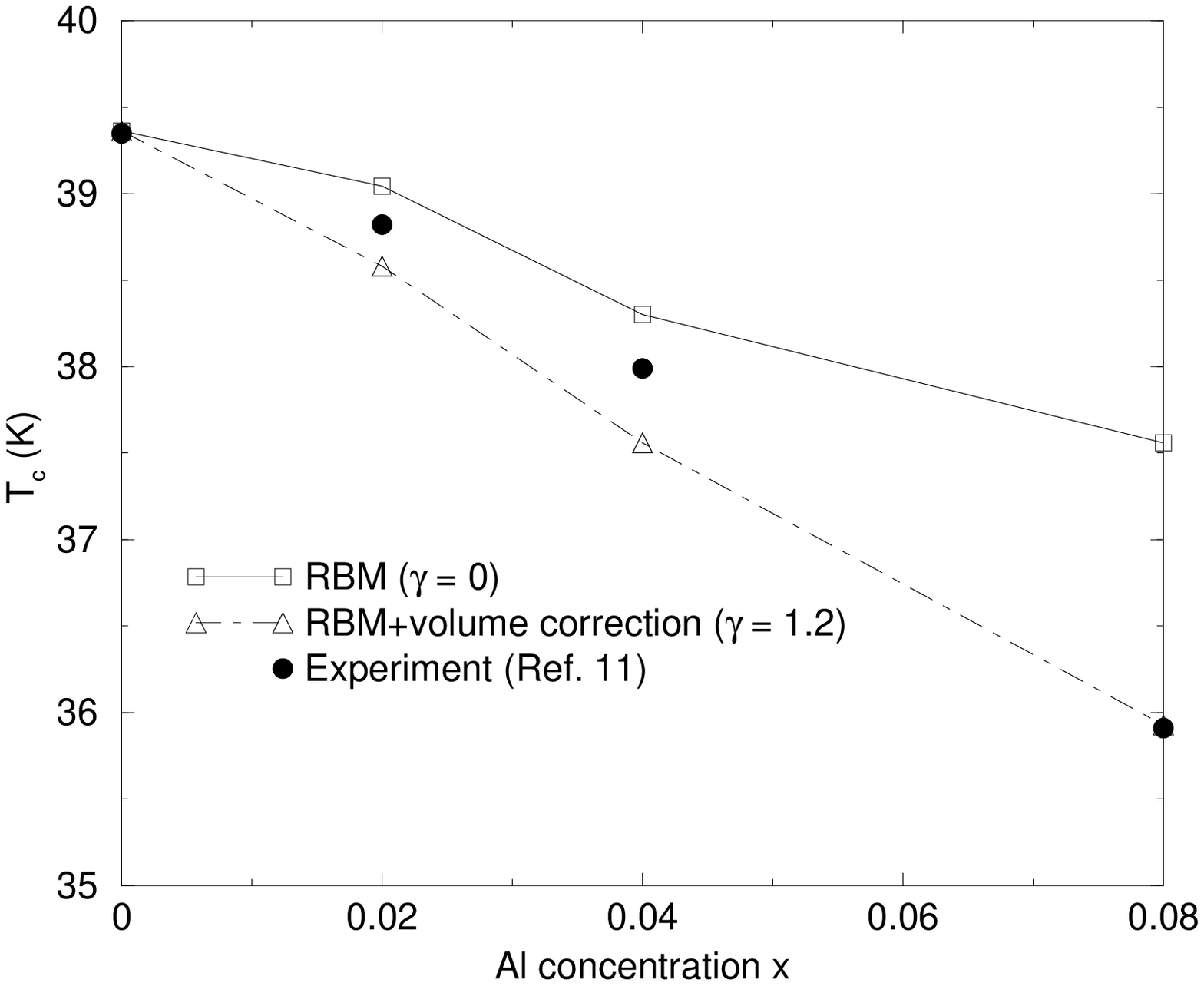, width=15cm}
\caption{Comparison of measured transition temperatures 
of Mg$_{1-x}$Al$_x$B$_2$ as a function of Al concentration ($x$) from Ref. \cite{Bianconi}
with those calculated from DOS obtained from a rigid band model (RBM) (open squares)
and from the parameters in Table III with corrections for the increase in volume 
(``RBM+volume correction'') (open triangles). The lines simply serve to guide the eye.}
\label{tcab3}
\end{figure}


\begin{references}
\bibitem{note} Note that the McMillan approximation to the full Eliashberg
equations is highly accurate for all superconductors with
$\lambda<1.5$ once a proper average prefactor 
is considered, as shown in Ref.\cite{Allen}.
\end{references}
\end{document}